# *Enhancing the Directional Violation of Kirchhoff's Law of Thermal Radiation with a Nonreciprocal Wire Medium*

David E. Fernandes[1], Mário G. Silveirinha[1,2]

[1] *Instituto de Telecomunicações, Avenida Rovisco Pais, 1, 1049-001 Lisboa, Portugal l*

[2] *University of Lisbon, Instituto Superior Técnico, Avenida Rovisco Pais, 1, 1049-001 Lisboa, Portugal*

**Abstract**

In this work, we develop a homogenization model to determine the effective response of a metallic nanowire array embedded in an electric gyrotropic material. We study the interaction of electromagnetic waves with the metamaterial and demonstrate that the nanowire array can greatly enhance the nonreciprocal response of the gyrotropic substrate. In particular, the metamaterial can either absorb the incoming energy almost entirely or reflect it with little loss, depending on the sign of the incidence angle. We explore the implications of our findings in the context of Kirchhoff's law of thermal radiation. Our results demonstrate that the wire array can boost the difference between the emissivity and absorptivity in a broad spectrum of frequencies and incidence angles as compared to an unstructured gyrotropic substrate. These findings suggest potential applications for the nonreciprocal wire medium in thermal management, radiative cooling, and others.

[1] E-mail: dfernandes@co.it.pt

[2] To whom correspondence should be addressed: E-mail: mario.silveirinha@tecnico.ulisboa.pt

# I. Introduction

In recent years, there has been a renewed interest in nonreciprocal effects in photonic systems. Nonreciprocal media may enable unique wave phenomena, the most notable of which being the unidirectional propagation of light [1-8]. Electromagnetic reciprocity is rooted in three properties: linearity, conservation of energy and time-reversal symmetry [9]. The most common way to obtain a nonreciprocal response is by breaking the time-reversal symmetry of the photonic system with a suitable bias [10]. Typically, this is achieved by using materials with a gyromagnetic response, such as ferrites under a magnetic bias [11-14], but other alternatives have been discussed in the literature [15-27]. Interestingly, it has been shown that a nonreciprocal response has often a topological origin [28-30].

Another way to control and tailor the light wave propagation is by using artificial structured media. In particular, wire metamaterials have been extensively studied over the last two decades due their unusual electromagnetic properties, such as hyperbolic dispersion [31], high density of photonic states [32, 33], amongst others [34]. The standard wire medium consists of a set of long parallel metallic wires embedded in a dielectric material [35, 36]. Metamaterials formed by wire arrays have been widely used in the microwave through mid-THz frequency band [37-45].

In this work we bridge these two fields of research and study the response of a nonreciprocal wire metamaterial formed by metallic wires embedded in an electric-gyrotropic material. We introduce a homogenization model that accounts for scenarios involving an electric-gyrotropic host medium, thus expanding upon prior theories limited to an isotropic and reciprocal dielectric host [46-48]. Using the effective medium theory and full-wave numerical simulations we prove that the wave interactions with the metallic wires and the gyrotropic substrate lead to unique effects such as the formation of transparency bands in a frequency range wherein both the metal and the gyrotropic materials are opaque [49]. As shown in Ref. [49], the origin of the transparency bands is the formation of topological localized edge states at the interfaces of the metallic wires. Here, we study this phenomenon using a rigorous effective medium theory that describes the macroscopic metamaterial response that emerges from the hybridization of such localized states.

We apply our effective medium theory to investigate how the metallic wire array can tailor the scattering from a gyrotropic material slab. As is well known, the magnetic bias of the substrate can break some mirror symmetries (e.g., the $x$-parity symmetry $(x,y,z) \xrightarrow{\mathcal{P}_x} (-x,y,z)$), resulting in scattering parameters that are not invariant under a parity transformation of the incidence angle ($\theta \to -\theta$). For example, in our system we find that due to the magnetic bias the reflection coefficient is sensitive to the parity of the incidence angle $|R(\theta)| \neq |R(-\theta)|$. This asymmetry has important consequences in the context of thermal radiation. For reciprocal systems, Kirchhoff's law [50] implies that a material's angular and frequency-dependent emissivity ($e(\theta,\omega)$) and absorptivity ($\alpha(\theta,\omega)$) are identical ($\alpha(\theta,\omega) = \alpha(-\theta,\omega) = e(\theta,\omega) = e(-\theta,\omega)$). This property guarantees that in thermal equilibrium the radiation absorbed by a given body is identical to the radiation emitted by the same body. Interestingly, nonreciprocity leads to the weaker form of Kirchhoff's directional law, $\alpha(\theta,\omega) = e(-\theta,\omega)$, which still ensures that the total absorption (integrated over all directions of space) is identical to the total emission, even though locally (for a fixed direction of space) there is a violation of the detailed power balance: $\alpha(\theta,\omega) \neq e(\theta,\omega)$. In recent years, different groups have shown that systems with a nonreciprocal response may be used to break the directional power balance so that there is a significant imbalance between the emissivity and absorptivity spectra [51-56]. The directional violation of Kirchhoff's law has relevant applications in the context of thermal management, including the design of ideal solar absorbers, cooling devices with zero absorptivity and perfect emissivity [51], amongst others. Here, we show that loading a gyrotropic nonreciprocal material with the metallic nanowires can greatly enhance the asymmetry between the absorptivity and the emissivity across a wide range of incidence angles.

This paper is organized as follows. In Sec. II we introduce an effective medium model for a wire-medium embedded in a gyrotropic material and study the bulk plane-wave modes in the metamaterial. It is shown that the interactions between the metallic wires and gyrotropic host lead to the formation of transparency bands. In Sec. III we investigate the scattering of light by a metamaterial slab, demonstrating that the reflection level is strongly dependent on the sign of the incidence angle. We discuss the implications of this effect in the context of Kirchhoff's law of thermal radiation. Finally, in Sec. IV the conclusions are drawn. The time dependence of the electromagnetic fields is of the type $e^{-i\omega t}$.

## II. Homogenization model

### A. Effective permittivity

Here we determine the effective permittivity of a metamaterial formed by an array of metallic wires with radius $r_w$ arranged in a square lattice with period $a$. The metallic wires have permittivity $\varepsilon_0 \varepsilon_m$ and are embedded in gyrotropic host material with permittivity $\bar{\varepsilon}_h$. The response of the host material is tailored by a static magnetic field $\mathbf{B}_0 = +B_z \hat{\mathbf{z}}$ oriented along +z-direction. The relative permittivity tensor of the host medium is:

$$\frac{\bar{\varepsilon}_h}{\varepsilon_0} = \varepsilon_t \mathbf{1}_t + i\varepsilon_g \hat{\mathbf{z}} \times \mathbf{1}_t + \varepsilon_a \hat{\mathbf{z}} \otimes \hat{\mathbf{z}} = \begin{pmatrix} \varepsilon_t & -i\varepsilon_g & 0 \\ i\varepsilon_g & \varepsilon_t & 0 \\ 0 & 0 & \varepsilon_a \end{pmatrix}, \quad (1)$$

with $\mathbf{1}_t = \mathbf{1} - \hat{\mathbf{z}} \otimes \hat{\mathbf{z}}$, where $\mathbf{1}$ is the identity matrix. For a magnetized electric plasma the components of the permittivity tensor are:

$$\varepsilon_t = 1 - \frac{\omega_p^2(\omega+i\Gamma)}{\omega(\omega+i\Gamma)^2 - \omega\omega_0^2}, \quad \varepsilon_g = \frac{\omega_p^2 \omega_0}{\omega\omega_0^2 - \omega(\omega+i\Gamma)^2}, \quad \varepsilon_a = 1 - \frac{\omega_p^2}{\omega(\omega+i\Gamma)}, \quad (2)$$

with $\omega_p$ the plasma angular frequency, $\omega_0 = -qB_z/m^* = +eB_z/m^*$ the cyclotron frequency and $\Gamma$ is the collision frequency associated with material loss. For example, the metamaterial may be formed by cylindrical rods of a noble metal (e.g., silver) embedded in a magnetized semiconductor (e.g., InSb). Note that we neglect the effect of the bias magnetic field on the metal response, and hence the permittivity of the rods is a scalar. This approximation is justified because the plasma frequency of the metal is several orders of magnitude larger than the cyclotron frequency.

In Appendix A, we derive the effective permittivity $\bar{\bar{\varepsilon}}_{ef}$ of the considered gyrotropic wire metamaterial. In the long wavelength regime ($a \ll \lambda$ and $ka \ll \pi$) and for thin rods ($r_w \ll a$) oriented along the $y$-direction, one obtains

$$\frac{\bar{\bar{\varepsilon}}_{ef}}{\varepsilon_0} = \frac{\bar{\bar{\varepsilon}}_h}{\varepsilon_0} + \frac{\hat{\mathbf{y}} \otimes \hat{\mathbf{y}}}{\frac{\varepsilon_m - \varepsilon_t}{(\varepsilon_m - \varepsilon_t)^2 - \varepsilon_g^2} \frac{1}{f_V} - \left( \frac{1}{\beta_p^2} \left(\frac{\omega}{c}\right)^2 - \frac{1}{\beta_\varepsilon^2(\omega)} k_y^2 \right)}, \quad (3)$$

where $f_V = \pi r_w^2 / a^2$ is the volume fraction of the metal. The second term in the right-hand side describes the homogenized response of the metallic rods in the gyrotropic host background. This term is a function of both frequency $\omega$ and of the wave vector component $k_y$ along the wires. The response of the rods depends on two parameters $\beta_p$ and $\beta_\varepsilon(\omega)$ defined as:

$$\frac{1}{\beta_p^2} = \left(\frac{a}{2\pi}\right)^2 \sum_{(m,n) \neq (0,0)} \frac{\left[J_0\left(\sqrt{m^2+n^2}\,\frac{2\pi r_w}{a}\right)\right]^2}{m^2+n^2}, \quad (4a)$$

$$\frac{1}{\beta_\varepsilon^2(\omega)} = \left(\frac{a}{2\pi}\right)^2 \sum_{(m,n) \neq (0,0)} \frac{\left[J_0\left(\sqrt{m^2+n^2}\,\frac{2\pi r_w}{a}\right)\right]^2}{\varepsilon_t(\omega)m^2 + \varepsilon_a(\omega)n^2}. \quad (4b)$$

In the above, $J_0$ is the Bessel function of first kind and order zero.

It is useful to note that for a regular isotropic reciprocal host with $\bar{\varepsilon}_h = \varepsilon_h \mathbf{1}$, one has $\varepsilon_t(\omega) = \varepsilon_a(\omega) \to \varepsilon_h$ and $\varepsilon_g(\omega) = 0$, so that $\beta_\varepsilon^2(\omega) = \varepsilon_h \beta_p^2$. In that case, the homogenization model reduces to that of Ref. [47]:

$$\frac{\bar{\varepsilon}_{ef}}{\varepsilon_0} = \varepsilon_h \left[ \mathbf{1} + \frac{\hat{\mathbf{y}} \otimes \hat{\mathbf{y}}}{\dfrac{1}{\varepsilon_m/\varepsilon_h - 1}\dfrac{1}{f_V} - \dfrac{1}{\beta_p^2}\left(\varepsilon_h\left(\dfrac{\omega}{c}\right)^2 - k_y^2\right)} \right], \quad \text{(isotropic host).} \quad (5)$$

This observation shows that the parameter $\beta_p$ can be identified with the plasma wave number of an array of perfect electrical conducting (PEC) rods [46, 47]:

$$\beta_p \approx \frac{1}{a}\sqrt{\frac{2\pi}{\ln\left(\dfrac{a}{2\pi r_w}\right) + 0.5275}}. \quad (6)$$

It can be numerically checked that Eqs. (4a) and Eq. (6) yield very similar results for $r_w/a < 0.1$. The difference is typically less than 1%.

For simplicity, in this study we assume that the metallic rods can be modeled as perfect conductors so that $\varepsilon_m \to -\infty$. This is a good approximation provided the radius of the rods exceeds a few times the skin depth of the metal [59], which is typically the case at THz frequencies, i.e. near the plasma frequency of the semiconductor. In such a case, the effective permittivity model (3) reduces to:

$$\frac{\bar{\varepsilon}_{ef}}{\varepsilon_0} = \frac{\bar{\varepsilon}_h}{\varepsilon_0} - \beta_p^2 \frac{\hat{\mathbf{y}} \otimes \hat{\mathbf{y}}}{\left(\dfrac{\omega}{c}\right)^2 - \dfrac{\beta_p^2}{\beta_\varepsilon^2(\omega)} k_y^2}. \quad (7)$$

## B. Quasi-static resonances

For a gyrotropic host, the parameter $\beta_\varepsilon(\omega)$ depends on a complicated manner on the host material permittivity and cannot be expressed in terms of $\beta_p$. Interestingly, in the lossless limit ($\Gamma \to 0$), $\varepsilon_t(\omega)$ and $\varepsilon_a(\omega)$ are real-valued and can have opposite signs. Consequently, in the lossless limit $\beta_\varepsilon(\omega)$ can exhibit a resonant behavior due to singularities associated with the zeros of $\varepsilon_t(\omega)m^2 + \varepsilon_a(\omega)n^2 = 0$. As discussed next, these singularities have a quasi-static origin and are associated with "volume plasmons" of the bulk gyrotropic material.

Indeed, in the quasi-static limit one can write the electric field in terms of a scalar potential $\mathbf{E} \approx -\nabla\phi$ that satisfies $\nabla \cdot [\overline{\varepsilon}_h \cdot \mathbf{E}] = 0$. Considering plane wave solutions with a space-time variation of the form $e^{i\mathbf{k}\cdot\mathbf{r}} e^{-i\omega t}$, one readily obtains the dispersion relation $\mathbf{k} \cdot \overline{\varepsilon}_h \cdot \mathbf{k} = 0$, which is equivalent to $\varepsilon_t(k_x^2 + k_y^2) + \varepsilon_a k_z^2 = 0$. The solutions of $\varepsilon_t(k_x^2 + k_y^2) + \varepsilon_a k_z^2 = 0$ are the volume plasmons in the bulk gyrotropic material. Consider now that the array of metallic rods directed along the *y*-direction is embedded into the gyrotropic material. In the long wavelength limit, the interactions between the rods are mediated by wave vectors of the type $\mathbf{k} = \frac{2\pi}{a}(m\hat{\mathbf{x}} + n\hat{\mathbf{z}})$, with $m, n$ integers, i.e., are mediated by the reciprocal primitive vectors of the 2D lattice. The coupling with the volume plasmons leads to singularities at frequencies that satisfy $\varepsilon_t(\omega)m^2 + \varepsilon_a(\omega)n^2 = 0$, justifying the previously discussed resonant behavior of $\beta_\varepsilon(\omega)$. It can be shown that the range of frequencies associated with a resonant behavior is determined by $\omega \in [0, \omega_{\min}] \cup [\omega_{\max}, \sqrt{\omega_0^2 + \omega_p^2}]$ with $\omega_{\min} = \min\{|\omega_0|, \omega_p\}$ and $\omega_{\max} = \max\{|\omega_0|, \omega_p\}$. When dissipation is taken into account ($\Gamma > 0$) the resonant behavior is smeared out.

For instance, let us consider a gyrotropic medium characterized by $\omega_p/2\pi = 5$ THz, $\omega_0 = 0.5\omega_p$ and $\Gamma = 0.107\omega_p$, consistent with the loss level of n-doped InSb [57, 58]. Except if explicitly stated otherwise, these parameters are fixed from hereon. Figure 1 depicts the parameter $\beta_\varepsilon = \beta'_\varepsilon + i\beta''_\varepsilon$ calculated as function of the normalized frequency. The lattice period is fixed such that $a = c/\omega_p = 9.55$μm and we consider two sets of wires, one with radius $r_w = 0.05a = 477$nm (blue solid curves) and another with $r_w = 0.1a = 955$nm (black solid curves). The series (4b) is truncated to the range $|m|,|n| \leq 50$ in most of the examples of the article. The truncation limit guarantees numerical convergence for the level of loss of *n*-doped InSb.

Figure 1 reveals that for small frequencies $\beta_\varepsilon(\omega)$ (solid curves) can be much larger than the parameter $\beta_p$ (dashed curves). A large $\beta_\varepsilon(\omega)$ can be helpful to suppress the effects of spatial dispersion of the wire medium, i.e., to reduce the impact of the term that depends on $k_y$ in Eq. (3). The imaginary value of $\beta_\varepsilon(\omega)$ implies enhanced material absorption due to the interaction with the metallic wires. The two local peaks in $\beta''_\varepsilon$ are a residual effect of the quasi-static resonances discussed in the beginning of this subsection. For smaller values of $\Gamma$ the number and the strength of the resonances is enhanced (not shown). For frequencies $\omega \gg \omega_p$, the response of the gyrotropic medium becomes asymptotically similar to that of air and one obtains $\beta_\varepsilon(\omega) \approx \beta_p$.

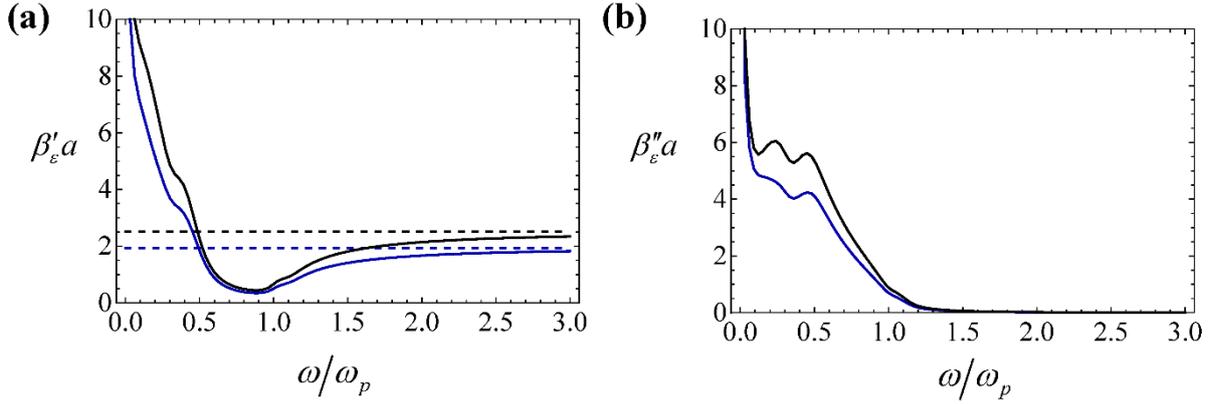

**Fig. 1 a)** Real and **b)** Imaginary parts of $\beta_\varepsilon = \beta'_\varepsilon + i\beta''_\varepsilon$ as a function of the normalized frequency. The host gyrotropic material is characterized by $\omega_0 = 0.5\omega_p$ and $\Gamma = 0.107\omega_p$. The metal wires have radius $r_w = 0.05a$ (blue solid curves) or $r_w = 0.1a$ (black solid curves). The dashed curves in panel a) represent the (frequency-independent) parameter $\beta_p$ for the two radii.

In general, the nonreciprocal effects are especially strong when $\omega_0$ and $\omega_p$ have comparable magnitude. Even though challenging such a scenario is realistic. The cyclotron frequency $\omega_0 = B_z e/m^*$ depends on the bias field $B_z$ and on the electron effective mass $m^*$. For narrow-gap semiconductors, such as InSb, the effective mass $m^*$ is small ($m^*_{InSb} = 0.022 m_0$ [57]) and thereby the cyclotron frequency can have a value on the order of a few THz. The required magnetic bias for $\omega_0 = 0.5\omega_p = 2\pi \times 2.5\,\text{THz}$ is about 2T.

## C. Plane wave modes and isofrequency surfaces

In order to understand the impact of the magnetic bias on the response of the metamaterial, first we characterize the dispersion of the bulk plane-wave modes. For simplicity, we restrict our attention to the wave propagation in the *xoy* plane, i.e., to the plane perpendicular to the magnetic bias. Moreover, we consider only TM-polarized waves with $\mathbf{H} = H_z(x,y)\hat{\mathbf{z}}$ and $\mathbf{E} = E_x(x,y)\hat{\mathbf{x}} + E_y(x,y)\hat{\mathbf{y}}$.

Substituting Eq. (7) into the Maxwell's equations and assuming a spatial variation of the type $e^{i\mathbf{k}\cdot\mathbf{r}} = e^{i(k_x x + k_y y)}$, one readily finds that the plane wave modes are constrained by:

$$\begin{bmatrix} k_y & -k_x & \mu_0\omega \\ \omega\varepsilon_0\varepsilon_{xx} & \omega\varepsilon_0\varepsilon_{xy} & k_y \\ \omega\varepsilon_0\varepsilon_{yx} & \omega\varepsilon_0\varepsilon_{yy} & -k_x \end{bmatrix} \cdot \begin{bmatrix} E_x \\ E_y \\ H_z \end{bmatrix} = 0, \tag{8}$$

with $\varepsilon_{ij}$ and $i, j = x, y$ the components of effective dielectric function $\bar{\bar{\varepsilon}}_{\text{ef}} / \varepsilon_0$ in Eq. (7), i.e.,

$$\varepsilon_{xx} = \varepsilon_t, \qquad \varepsilon_{xy} = -\varepsilon_{yx} = -i\varepsilon_g, \qquad \varepsilon_{yy} = \varepsilon_t - \frac{\beta_p^2}{\left(\frac{\omega}{c}\right)^2 - \frac{\beta_p^2}{\beta_\varepsilon^2(\omega)}k_y^2}. \tag{9}$$

The dispersion of the plane waves is found by setting the determinant of the matrix in Eq. (8) equal to zero, which yields:

$$k_x^2 \varepsilon_{xx} + k_y^2 \varepsilon_{yy} = \left(\varepsilon_{xx}\varepsilon_{yy} - \varepsilon_g^2\right)\left(\frac{\omega}{c}\right)^2. \tag{10}$$

The effect of metallic rods in the metamaterial response is fully described by the permittivity component $\varepsilon_{yy}$. Substituting Eq. (9) into Eq. (10) one obtains a second-degree equation for $k_y^2$ whose solutions are:

$$k_y^2 = \frac{-\beta_p^2\left(\beta_\varepsilon^2 + k_x^2\varepsilon_t\right) + \left(\frac{\omega}{c}\right)^2\left(\beta_\varepsilon^2\varepsilon_t - \beta_p^2\left(\varepsilon_t^2 - \varepsilon_g^2\right)\right) \pm B_1}{2\beta_p^2\varepsilon_t}, \text{ with} \tag{11a}$$

$$B_1 = \sqrt{4\left(\frac{\omega}{c}\right)^2 \beta_\varepsilon^2 \beta_p^2 \varepsilon_t \left(\left(k_x^2 + \beta_p^2\right)\varepsilon_t - \left(\frac{\omega}{c}\right)^2\left(\varepsilon_t^2 - \varepsilon_g^2\right)\right) + \left(\beta_p^2\left(\beta_\varepsilon^2 + k_x^2\varepsilon_t\right) - \left(\frac{\omega}{c}\right)^2\left(\beta_\varepsilon^2\varepsilon_t + \beta_p^2\left(\varepsilon_t^2 - \varepsilon_g^2\right)\right)\right)^2}$$
(11b)

Similar to the standard wire medium configuration, due to spatial dispersion the metamaterial supports two bulk TM modes. This is in contrast to the gyrotropic host medium, which supports only a single bulk TM mode.

Importantly, while both the bulk gyrotropic material and the metal do not support wave propagation for low frequencies, ($\text{Im}\{k_y\} \neq 0$ even in the lossless limit), a mixture of the two materials can create a "topological" transparency band that enables the wave propagation in the metamaterial in the long wavelength limit [49]. In fact, the gyrotropic material does not support bulk TM eigenmodes for frequencies smaller than $\omega < -\frac{\omega_0}{2} + \sqrt{\left(\frac{\omega_0}{2}\right)^2 + \omega_p^2}$, which for $\omega_0 = 0.5\omega_p$ corresponds to $\omega < 0.78\omega_p$. In contrast, as further discussed below, the metamaterial supports a bulk propagating mode (without a low-frequency cut-off) in the lossless limit.

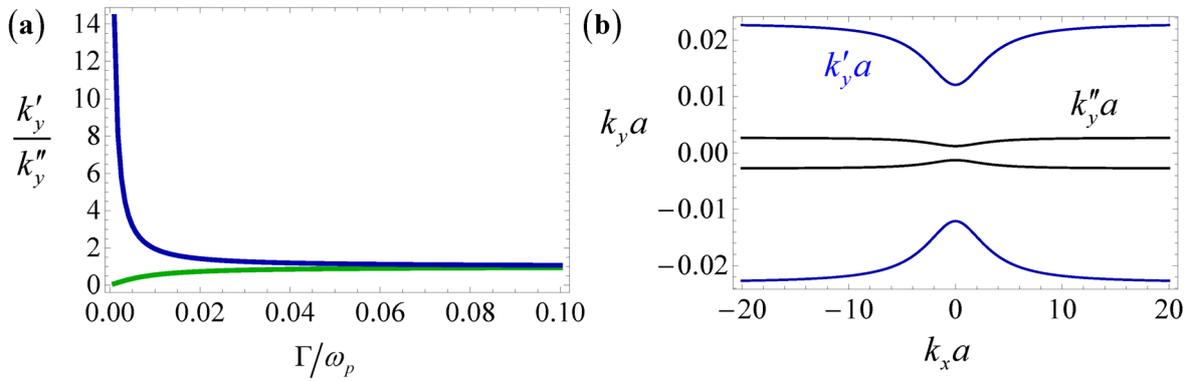

**Fig. 2 a)** Ratio between the real and imaginary parts of the propagation constant of the low-frequency bulk TM wave $k'_y/k''_y$ as a function of the normalized collision frequency for a fixed frequency of operation $\omega a/c = 0.005$ and $r_w = 0.05a$. Blue line: metamaterial [Eq. (11); Green line: bare gyrotropic host. **c)** Isofrequency contours of the wire-medium embedded in a gyrotropic medium $k_y = k'_y + ik''_y$ as a function of $k_x$ for $\omega a/c = 0.005$ and $\Gamma = 0.0015\omega_p$.

To demonstrate this property, we depict in Fig. 2a $k'_y/k''_y$, calculated for the metamaterial (blue line) and for the bare gyrotropic host (green line) as a function of the collision frequency. The frequency of operation is $\omega a/c = 0.005$ and $k_x = 0$. The radius of the metallic wires is $r_w = 0.05a = 477\text{nm}$. As seen, the host material is opaque due to a band-gap ($k''_y \gg k'_y$) even when the dissipation is extremely weak (green line). In contrast, the metamaterial can support a quasi-propagating mode with $k''_y \ll k'_y$ (blue curve). The topological mechanism at the origin of the transparency band in the metamaterial was discussed in Ref. [49].

Figure 2b shows the isofrequency contours for $\Gamma = 0.0015\omega_p$. As seen, similar to the dispersion of the quasi-TEM mode in the standard wire medium [47, 48], the isofrequency contours are nearly flat for large values of $k_x$. Thereby, notwithstanding the host material is opaque when on its own, the wire array may enable some form of wave "canalization" through the nonreciprocal gyrotropic host [49], even though it may be affected by the damping of the host material.

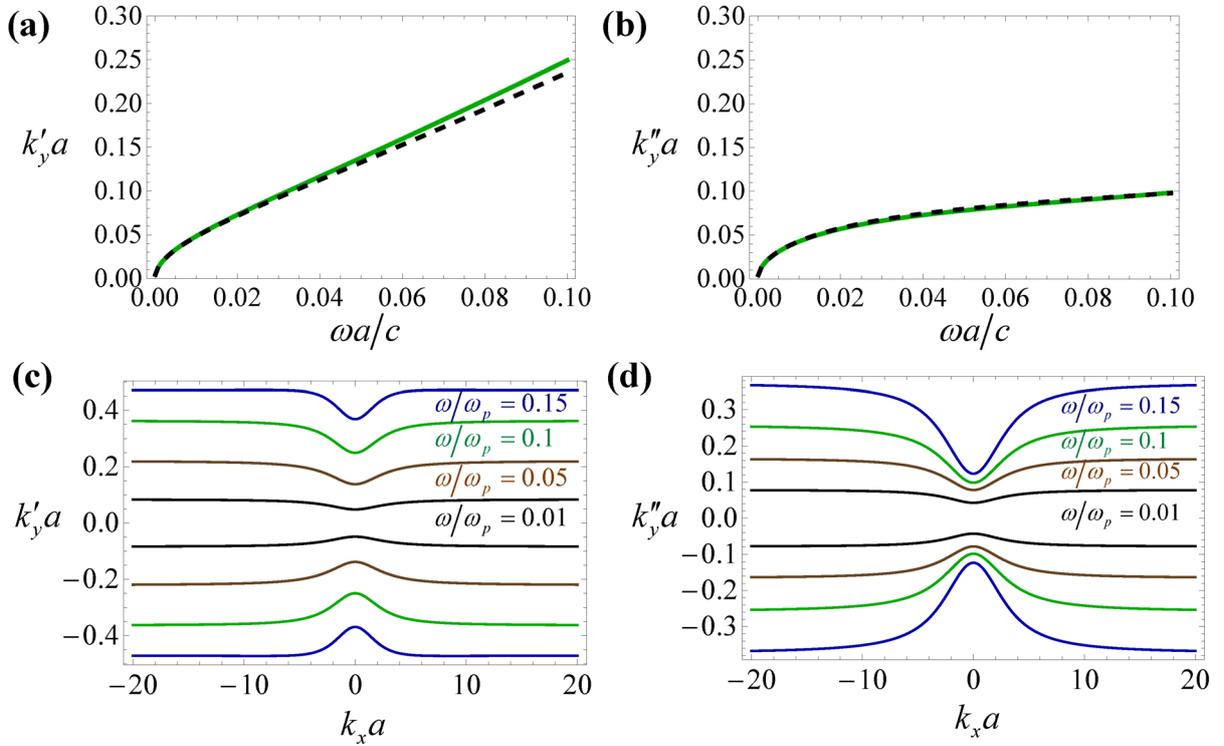

**Fig. 3 a)** Real and **b)** imaginary parts of the propagation constant $k_y = k'_y + ik''_y$ of the low-frequency bulk TM wave (solid green line) as a function of the normalized frequency. The metallic rods have radius $r_w = 0.05a$ and are embedded in the gyrotropic material. The dashed black curve depicts the relation $k_y = \sqrt{\varepsilon_t(\omega)}\,\omega/c$. **c)** and **d)** Isofrequency contours of the metamaterial $k_y = k'_y + ik''_y$ as a function of $k_x$ for different values $\omega/\omega_p$.

We show in Figs. 3a-b the dispersion of the quasi-propagating mode for InSb. As seen, for a realistic level of loss the attenuation constant $k''_y$ is comparable to the propagation constant $k'_y$. The bulk mode has a group velocity $v_{g,y} = \partial\omega/\partial k_y$ that is almost linear. Moreover, the dispersion follows closely the relation $k_y = \sqrt{\varepsilon_t(\omega)}\,\omega/c$ (dashed black curves in Fig.3a-b). Note that for low frequencies $\omega << |\omega_0|$, the transverse permittivity of the gyrotropic host satisfies $\varepsilon_t = 1 + \dfrac{\omega_p^2}{\omega_0^2}(1 + i\Gamma/\omega)$. Figures 3c-d depict the isofrequency contours of the quasi-TEM mode for several operating frequencies. As seen, despite the dissipative effects the isofrequency contours remain flat, analogous to the weak dissipation case discussed previously.

The other bulk TM mode of the gyrotropic wire medium is strongly attenuated for low frequencies ($k''_y \gg k'_y$). This can be seen in Fig. 4a, which depicts the dispersion relation of calculated using Eq. (11). Figures 4b-c show the corresponding isofrequency contours for a set of fixed frequencies of operation. The attenuation constant $k''_y$ increases with $|k_x|$. The imaginary part of the propagation constant of this mode dominates up to $\omega a/c \sim 1.1$. Beyond this threshold frequency, the mode exhibits propagative-type characteristics with $k'_y > k''_y$.

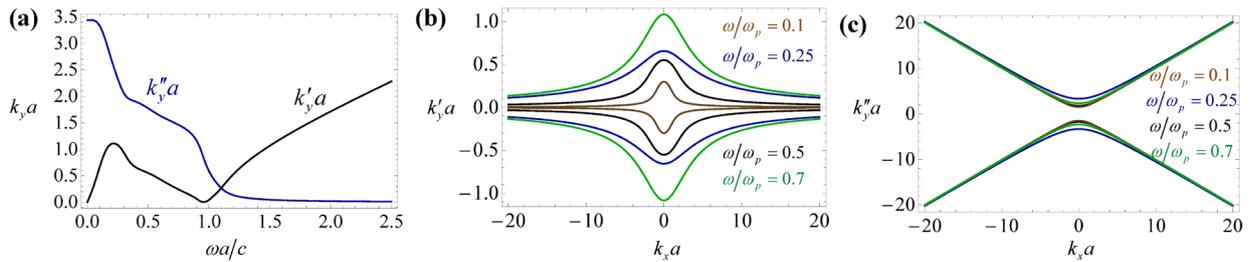

**Fig. 4 a)** Real and imaginary parts of the propagation constant of the second bulk TM wave $k_y = k'_y + ik''_y$ as a function of the normalized frequency for an array of rods with radius $r_w = 0.05a$ embedded in the gyrotropic material. **b)** and **c)** Isofrequency contours showing $k_y = k'_y + ik''_y$ as a function of $k_x$ for different values $\omega/\omega_p$.

## III. Asymmetric scattering and violation of the directional Kirchhoff's law

In what follows, we apply the effective medium theory developed in Section II to characterize the scattering of electromagnetic waves by a metamaterial slab with thickness $h$. A representative system geometry is depicted in Fig. 5. We consider that the incoming plane wave is TM polarized ($\mathbf{H} = H_z \hat{\mathbf{z}}$) so that the plane of incidence is the *xoy* plane. The variation of the fields along the *x*-direction is of the type $e^{ik_x x}$ with $k_x = \sin\theta \frac{\omega}{c}$ determined by the angle of incidence $\theta$.

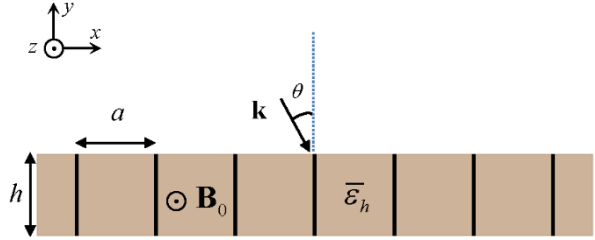

**Fig. 5** Geometry of the scattering problem under study. A gyrotropic wire medium slab with thickness $h$ is illuminated by a TM-polarized plane wave with an incidence angle $\theta$. The metallic wires are oriented along the *y*-direction, perpendicular to the interfaces. The host material is biased with static magnetic field $\mathbf{B}_0 = +B_z \hat{\mathbf{z}}$ oriented along the *z*-direction.

### A. *Analytical solution of the scattering problem*

We write the fields in the different regions of space as a superposition of plane waves. The magnetic field distribution in all space is determined by:

$$H_z = e^{ik_x x} \frac{E^{inc}}{\eta_0} \begin{cases} e^{-ik_{y0}y} - R e^{ik_{y0}y} & y > 0 \\ A_1^- e^{-ik_1(y+h)} + A_1^+ e^{ik_1(y+h)} + A_2^- e^{-ik_2(y+h)} + A_2^+ e^{ik_2(y+h)} & -h < y < 0, \\ T e^{-ik_{y0}(y+h)} & y < -h \end{cases} \quad (12)$$

where $\eta_0 = \sqrt{\mu_0/\varepsilon_0}$ is the free-space impedance, the constants $A_1^\pm, A_2^\pm$ are the amplitudes of the waves that propagate inside the slab, and $T$ and $R$ are the transmission and reflection coefficients, respectively. The propagation constant in the air regions is $k_{y0} = \sqrt{\left(\dfrac{\omega}{c}\right)^2 - k_x^2}$ and the propagation constants in the metamaterial $k_1$ and $k_2$ are determined by Eq. (11).

The electric field associated with each plane wave inside the metamaterial can be found using $\mathbf{E} = \dfrac{1}{-i\omega} \overline{\overline{\varepsilon}}_{ef}^{-1} \cdot \nabla \times \mathbf{H}$, which is the same as

$$\mathbf{E} = \frac{1}{-i\omega\varepsilon_0} \frac{1}{\varepsilon_{xx}\varepsilon_{yy} - \varepsilon_g^2} \left[ \left( \varepsilon_{yy}\partial_y H_z + \varepsilon_g k_x H_z \right)\hat{\mathbf{x}} + \left( -ik_x \varepsilon_{xx} H_z - i\varepsilon_g \partial_y H_z \right)\hat{\mathbf{y}} \right], \quad (13)$$

with $\partial_x = \partial/\partial x$ and $\partial_y = \partial/\partial y$. From here one obtains the electric field distribution:

$$E_x = \frac{e^{ik_x x}}{-i\omega\varepsilon_0} \frac{E^{\text{inc}}}{\eta_0} \begin{cases} -ik_0 \left( e^{-ik_{y0}y} + R e^{ik_{y0}y} \right) & y > 0 \\ A_1^- \Theta_1^- e^{-ik_1(y+h)} + A_1^+ \Theta_1^+ e^{ik_1(y+h)} + A_2^- \Theta_2^- e^{-ik_2(y+h)} + A_2^+ \Theta_2^+ e^{ik_2(y+h)} & -h < y < 0 \\ T\gamma_0 e^{-ik_{y0}(y+h)} & y < -h \end{cases} \quad (14a)$$

$$E_y = \frac{e^{ik_x x}}{-i\omega\varepsilon_0} \frac{E^{\text{inc}}}{\eta_0} \begin{cases} -ik_x \left( e^{-ik_{y0}y} - R e^{ik_{y0}y} \right), & y > 0 \\ A_1^- \Delta_1^- e^{-ik_1(y+h)} + A_1^+ \Delta_1^+ e^{ik_1(y+h)} + A_2^- \Delta_2^- e^{-ik_2(y+h)} + A_2^+ \Delta_2^+ e^{ik_2(y+h)}, & -h < y < 0, \\ -ik_x T e^{-ik_{y0}(y+h)}, & y < -h \end{cases} \quad (14b)$$

with $\Theta_m^\pm = \dfrac{(\pm ik_m)\varepsilon_{yy} + \varepsilon_g k_x}{\varepsilon_{xx}\varepsilon_{yy} - \varepsilon_g^2}$ and $\Delta_m^\pm = \dfrac{-ik_x \varepsilon_{xx} - i\varepsilon_g (\pm ik_m)}{\varepsilon_{xx}\varepsilon_{yy} - \varepsilon_g^2}$, for $(m = 1, 2)$. Due to the spatial dispersion effects, the permittivity $\varepsilon_{yy}$ depends on the propagation constant of each plane wave so that $\varepsilon_{yy} = \varepsilon_{yy}(\omega, k_m)$.

To determine the unknowns $A_1^\pm, A_2^\pm, R$ and $T$ it is necessary to impose suitable boundary conditions at the metamaterial-air interfaces. As usual, the tangential field components $H_z, E_x$ must be continuous at the interfaces. Furthermore, due to the strong spatial dispersion of the wire medium an additional boundary condition (ABC) is required [60-62]. The ABC imposes that the microscopic current $I$ flowing on the metallic wires vanishes at the slab interfaces. The contribution of the metallic wires to the macroscopic response may be expressed as $\mathbf{P}_\mathrm{w} = \left( \overline{\overline{\varepsilon}}_\mathrm{ef} - \overline{\overline{\varepsilon}}_h \right) \cdot \mathbf{E} = \varepsilon_0 \left( \varepsilon_{yy} - \varepsilon_t \right) \hat{\mathbf{y}}\hat{\mathbf{y}} \cdot \mathbf{E}$, where $\varepsilon_{yy} = \varepsilon_{yy}\left(\omega, k_y\right) = \varepsilon_{yy}\left(\omega, -i\partial_y\right)$. In order that $I$ vanishes, we must enforce that $\left[ \varepsilon_{yy}\left(\omega, -i\partial_y\right) - \varepsilon_t \right] E_y$ vanishes at the metamaterial side of the interfaces. In summary, the boundary conditions are:

$$\lfloor E_x \rfloor_{y=0,-h} = 0, \tag{15a}$$

$$\lfloor H_z \rfloor_{y=0,-h} = 0. \tag{15b}$$

$$\left[ \varepsilon_{yy}\left(\omega, -i\partial_y\right) - \varepsilon_t \right] E_y \bigg|_{y=0^-,-h^+} = 0 \tag{15c}$$

where $\lfloor F \rfloor_{y=y_0} = F_{y=y_0^+} - F_{y=y_0^-}$ stands for the field discontinuity of $F$ at the pertinent interface. These boundary conditions yield a linear system of equations that can be numerically solved to determine the reflection and transmission coefficients.

### B.    Transparency Bands

To begin with, we analyze the scattering of plane waves as a function of the operating frequency for the incidence angles $\theta = -60º$ and $\theta = -30º$. The metamaterial slab has thickness $h = 6a$. The remaining structural parameters are as in Sect. II.C. Figure 6 shows the reflection and transmission coefficients calculated using both the homogenization model (solid curves) and with the full-wave electromagnetic simulator CST Studio Suite [63] (discrete symbols).

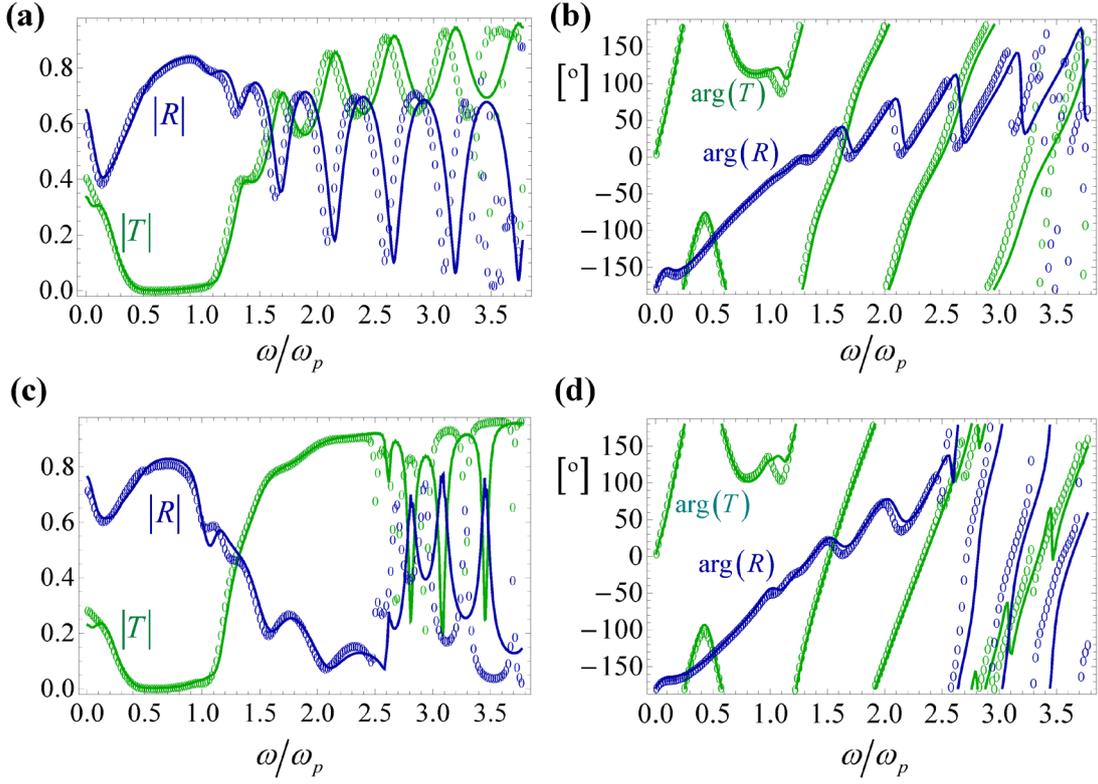

Fig. 6 **a)** Amplitude and **b)** Phase of the reflection (blue) and transmission (green) coefficients calculated as a function of frequency for a metamaterial slab with thickness $h = 6a$ at an incidence angle $\theta = -60º$. **c)** and **d)** are similar to **a)** and **b)** but for the incidence angle $\theta = -30º$. For all the panels, the solid curves are obtained with the homogenization model and the discrete symbols with the full wave simulations.

On the overall, there is quite good coincidence between the homogenization and full-wave results, which confirms the validity of the effective medium model. The agreement is especially good when $\omega a / c \ll 1$ so that the wavelength of operation is much longer than the lattice period. Curiously, for very small frequencies there is a small discrepancy between the effective medium model and the numerical simulations. The reason may be related to the fact that for the TM (non-propagative) mode $k_y a$ can be fairly large in the static-limit (see Fig. 4a). Hence, the Taylor approximation discussed in Appendix A (which requires $|k_y| a \ll 1$) may not be sufficiently accurate at very low frequencies for this particular mode. For large frequencies, the response of the gyrotropic host becomes similar to that of air, so that $\beta_\varepsilon \approx \beta_p$, and the metamaterial behaves as a standard wire medium with the metal rods embedded in air.

Interestingly, the transmission level is quite significant for small frequencies wherein the gyrotropic host is opaque. To better illustrate the role of the metallic wires, in Fig. 7a we compare the response of the metamaterial slab with the response of an unstructured gyrotropic material slab, for the same parameters as in Fig. 6. As seen, the transmission across a gyrotropic slab is vanishingly weak for the unstructured case (brown and black curves), whereas for the metamaterial slab (blue and green curves) it is at least one order of magnitude larger.

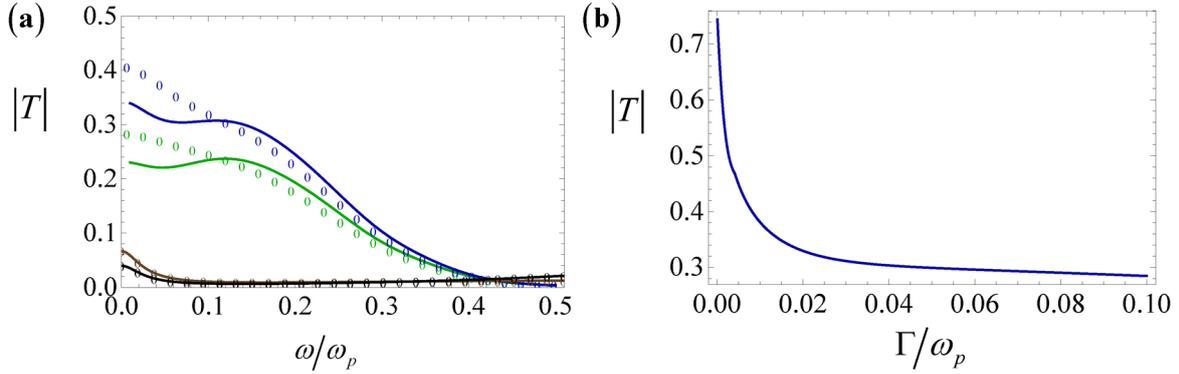

Fig. 7 **a)** Amplitude of the transmission coefficient calculated as a function of frequency for a metamaterial slab (blue and green curves, for $\theta = -60º$ and $\theta = -30º$) and for an unstructured gyrotropic material slab (brown and black curves, for $\theta = -60º$ and $\theta = -30º$). The system parameters are as in Fig. 6. The discrete symbols are calculated with the full wave simulator. **b)** Amplitude of the transmission coefficient as a function of the normalized collision frequency $\Gamma/\omega_p$ obtained with the full wave simulator.

The transmission level in the transparency window is limited by the significant level of loss of InSb. In Fig. 7b we show that the transmission amplitude can be greatly enhanced if the collision frequency of the host material is made smaller. For example, the transmission level can reach $|T| \approx 0.75$ when $\Gamma/\omega_p = 10^{-4}$. This high level of transmission may be attributed to the quasi-TEM propagation "channel" created by the wire array [49]. It should be noted that the transmission level may differ from unity in the lossless regime due to the effect of reflections.

In order to confirm the quasi-TEM nature mode of the modes propagating in the slab, we show in Fig. 8 the electromagnetic field distribution in all space for a frequency $\omega = 0.05\omega_p$ calculated using the full-wave simulator. [63]. As seen, apart from small perturbations near the metamaterial-air interfaces, both the electric and magnetic fields inside the slab are perpendicular to the wires, which is consistent with the properties of a quasi-TEM mode. The electric field in the region $y > 0$ is almost perpendicular to the interface due to the high-level of reflections.

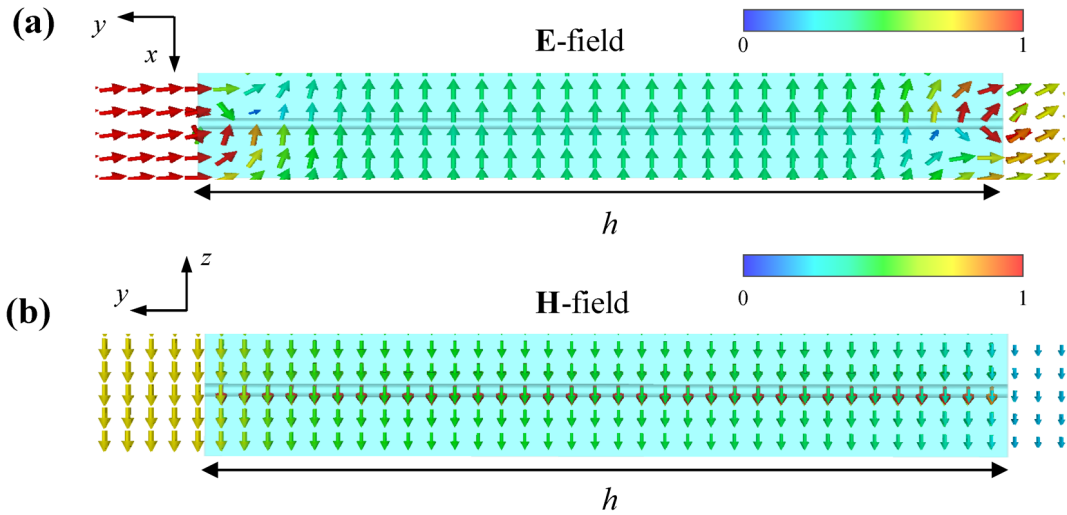

Fig. 8 **a)** Normalized electric field distribution in a unit cell of the metamaterial for the example of Fig. 6c-d at the fixed frequency $\omega = 0.05\omega_p$. **b)** Similar to **a)** but for the magnetic field.

### C.  *Violation of directional Kirchhoff's law*

Due to the nonreciprocity of the host material, the metamaterial scatters electromagnetic waves in an asymmetric manner. Indeed, plane waves illuminating the slab with incidence angles linked by parity, i.e. angles $\theta$ and $-\theta$, are scattered differently so that $R(\theta) \neq R(-\theta)$. Figure 9a depicts the amplitude of the reflection coefficient as a function of frequency for symmetric angles of incidence $\theta = 30º$ and $\theta = -30º$. As seen, for some frequencies there is a significant asymmetry in the reflection level, which is especially strong ($\left||R(-\theta)|-|R(\theta)|\right| \approx 0.5$ with $|R(\theta)| \approx 0$) near $\omega/\omega_p = 1.2$. In fact, for the considered geometry, the asymmetry is particularly strong in the frequency range $1.1 < \omega/\omega_p < 1.4$. Depending on the sign of the incidence angle, the metamaterial can exhibit either strong reflective properties or near-perfect absorption capabilities.

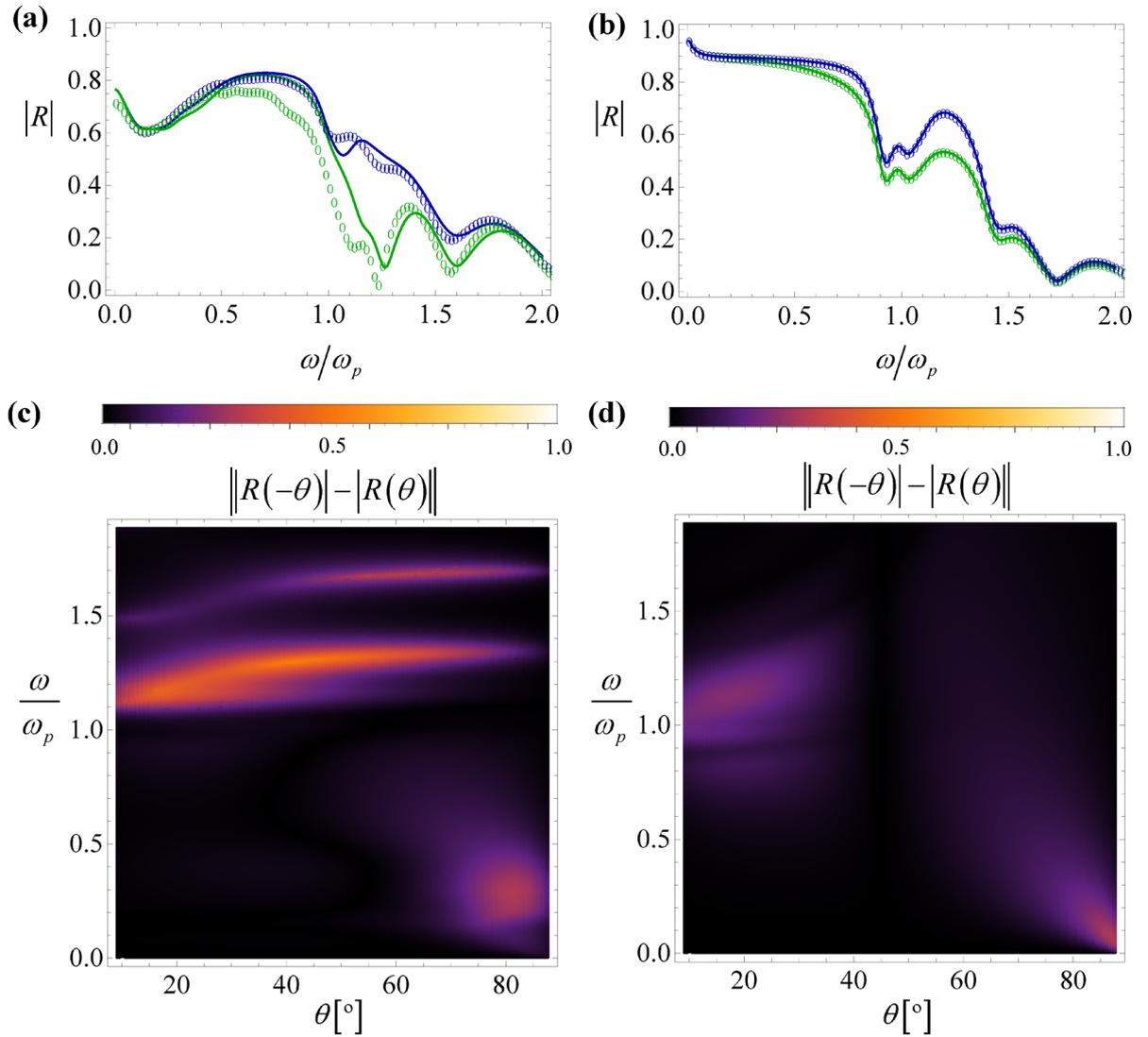

Fig. 9 **a)** Amplitude of the reflection coefficient of a metamaterial slab for the incidence angle $\theta = -30º$ (blue curve) and $\theta = 30º$ (green curve) calculated as a function of frequency. The solid curves are obtained with the homogenization model and the discrete symbols with the full wave simulations. **b)** Similar to **a)** but for a gyrotropic material slab without the wire array loading. **c)** Density plot of $\left||R(-\theta)|-|R(\theta)|\right|$ calculated using the homogenization theory as a function of the incidence angle and frequency. **d)** Similar to **c)** but for the gyrotropic slab.

Figure 9c presents a detailed study of the reflection asymmetry as a function of frequency and of the incidence angle. The reflection asymmetry is more pronounced slightly above the plasma frequency of the gyrotropic material over a wide range of incidence angles. In general, the scattering asymmetry can be enhanced either by using a stronger bias magnetic field or a thicker metamaterial slab. Thicker slabs enhance the nonreciprocity effects due to the stronger material absorption.

It is formally demonstrated in Appendix B, that due to a combination of spatial and time symmetries the transmission amplitude is independent of the sign of the incidence angle, i.e. $|T(-\theta)| = |T(\theta)|$. It should be noted that $|T(-\theta)| = |T(\theta)|$ also implies that $|R(-\theta)| = |R(\theta)|$ in a lossless system. Thus, material loss (ubiquitous in all materials) is mandatory to observe the discussed asymmetry.

Evidently, as the scattering asymmetry is rooted in the nonreciprocal response of the host material, it can occur even without the metallic wires. Interestingly, the asymmetry observed for the bare gyrotropic material slab is considerably weaker than in the composite material (see Figs. 9b and 9d). Hence, the scattering asymmetry is effectively enhanced by the wire medium loading. In other words, through the nanostructuring of the bare gyrotropic host it may be possible to boost the nonreciprocal effects. Since the reflection coefficients calculated at the top and bottom interfaces are related by $R_{\mathrm{BOT}}(\pm\theta) = R_{\mathrm{TOP}}(\mp\theta)$ (see Appendix B), our results can be readily generalized to the case of an illumination from the bottom side of the metamaterial slab.

The dissimilar scattering of waves may have interesting implications in thermal energy management. The Kirchhoff's law of thermal radiation establishes that in thermal equilibrium, the thermal emissivity *e* and absorptivity $\alpha$ for a certain direction of illumination satisfy [52, 53]:

$$\alpha(\theta,\omega) = 1 - \rho(\theta,\omega), \tag{16a}$$

$$e(\theta,\omega) = 1 - \rho(-\theta,\omega). \tag{16b}$$

where $\rho(\theta,\omega) = |R(\theta,\omega)|^2 + |T(\theta,\omega)|^2$ is determined by the scattering parameters for illumination along $\theta$. In reciprocal materials $R(\theta) = R(-\theta)$ and $T(\theta) = T(-\theta)$, and thereby in such a case the emissivity and absorptivity are identical. However, in nonreciprocal systems the detail balance $\eta(\theta,\omega) = |\alpha(\theta,\omega) - e(\theta,\omega)| = \left||R(\theta,\omega)|^2 - |R(-\theta,\omega)|^2\right|$ can be nontrivial for a wide range of incidence angles and operating frequencies. In the previous formula, we took into account that $T(\theta) = T(-\theta)$ (see Appendix B). Furthermore, for simplicity we restrict our discussion to directions $\theta$ in the *xoy* plane so that transverse electric (TE) polarized waves do not play any role in the calculation of $\eta(\theta,\omega)$.

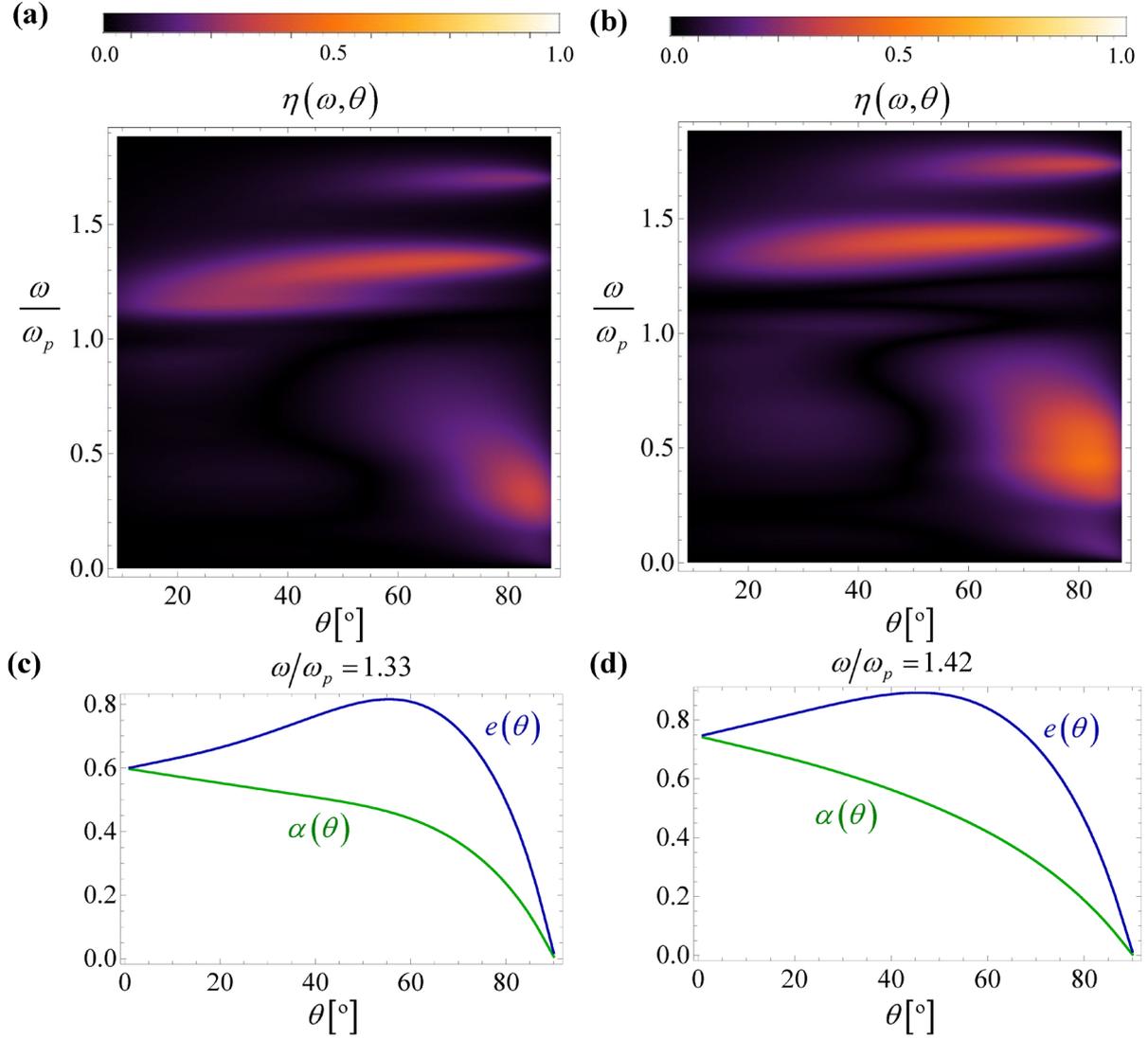

Fig. 10 **a)** Density plot of $\eta$ calculated using the homogenization theory as a function of the incidence angle and frequency for $\omega_0/\omega_p = 0.5$. **b)** Similar to **a)** but considering a gyrotropic host medium with $\omega_0/\omega_p = 0.75$ **c)** Emissivity $e$ (blue curve) and absorptivity $\alpha$ (green curve) for a fixed frequency of operation $\omega/\omega_p = 1.33$ calculated as a function of incidence angle for the case **a)**. **d)** Similar to **c)** but for the case **b)** and $\omega/\omega_p = 1.42$.

Figure 10a shows $\eta(\theta, \omega)$ calculated with our effective medium formalism as a function of frequency and incidence angle. As seen, the parameter $\eta$ can have a noteworthy value across a significant frequency range and for a multitude of incidence angles. The parameter $\eta$ is peaked near the frequency of operation $\omega/\omega_p = 1.33$, where the emissivity and absorptivity of the metamaterial slab differ by $e(\theta) - \alpha(\theta) \approx 0.35$ when $\theta \approx 64º$ (see Fig. 10c).

When the intensity of the bias field is raised to a level such that $\omega_0/\omega_p = 0.75$, the results for $\eta$ are qualitatively similar, apart from a slight blue shift in frequency (Fig. 10b). In fact, as shown in Fig. 10d, at the frequency where the peak of $\eta$ occurs ($\omega/\omega_p = 1.42$), the difference between the emissivity and absorptivity of the metamaterial is $e(\theta) - \alpha(\theta) \approx 0.4$. Given that the asymmetric reflectivity can cover a wide frequency range and a multitude of angles of incidence, we believe that the proposed metamaterial may be useful in realization of novel types of devices for thermal energy management.

## IV. Conclusions

In this work we developed a homogenization model for a wire medium embedded in a gyrotropic host substrate. It was shown using the effective medium model that the interactions between the metallic wires and the gyrotropic host can create a low-frequency propagative-type mode in the spectral range wherein the host medium has a complete bandgap. As an application of the theory, we studied the directional dependence of the reflectivity of the metamaterial. It was found that the wave scattering can be strongly asymmetric, such that the material may reflect strongly an incoming wave for some incidence angle, while efficiently absorbing an incident wave arriving from the parity-symmetric direction. Even though a related effect may be observed with a bare gyrotropic host, our theory and simulations demonstrate that the scattering asymmetry is dramatically enhanced with the metallic rod inclusions. The scattering asymmetry is especially pronounced near the plasma frequency of the host material, and most remarkably the metamaterial effect makes the asymmetry largely insensitive to the incidence angle. Finally, we highlighted that the metamaterial response enables a strong violation of the directional Kirchhoff's law of thermal radiation and thereby may have interesting applications in radiative cooling, thermophotovoltaics, and other related applications.

**Acknowledgements**

This work is supported in part by the Institution of Engineering and Technology (IET) and by Fundação para Ciência e a Tecnologia (FCT) under project UID/50008/2020. D. E. Fernandes acknowledges financial support by IT-Lisbon under the research contract with reference C-0042-22.

# Appendix A: The effective permittivity function

In the following, we derive an analytical formula for the effective permittivity of the metamaterial. To that end, first we generalize the homogenization theory of Ref. [64] to the case in which the host medium is a gyrotropic material. In a second step, we obtain an approximate analytical solution of the homogenization problem for the geometry of interest.

### A. The homogenization problem

To begin with, we consider the "microscopic" frequency dependent Maxwell equations in a non-magnetic medium:

$$\nabla \times \mathbf{E} = i\omega \mathbf{B} \tag{A1a}$$

$$\nabla \times \frac{\mathbf{B}}{\mu_0} = \mathbf{j}_e - i\omega \overline{\varepsilon}(\mathbf{r},\omega) \cdot \mathbf{E}. \tag{A1b}$$

Here, $\mathbf{E}, \mathbf{B}$ are the microscopic fields, $\omega$ is the oscillation frequency, and $\overline{\varepsilon}(\mathbf{r},\omega)$ is the position and frequency dependent microscopic permittivity of the composite medium. Furthermore, $\mathbf{j}_e = -i\omega \mathbf{P}_e e^{+i\mathbf{k}\cdot\mathbf{r}}$ represents a macroscopic current excitation characterized by the wave vector $\mathbf{k}$ and by the constant vector $\mathbf{P}_e$. For convenience we rewrite (A1b) as

$$\nabla \times \frac{\mathbf{B}}{\mu_0} = \mathbf{j}_e + \mathbf{j}_d - i\omega \overline{\varepsilon}_h \cdot \mathbf{E}, \tag{A1c}$$

where $\mathbf{j}_d = -i\omega(\overline{\varepsilon} - \overline{\varepsilon}_h) \cdot \mathbf{E}$ is the induced current relative to the gyrotropic host with permittivity $\overline{\varepsilon}_h$.

The macroscopic electromagnetic fields are obtained by averaging the microscopic fields as $\mathbf{E}_{av} = \frac{1}{V_{cell}} \int_\Omega \mathbf{E} e^{-i\mathbf{k}\cdot\mathbf{r}} d^3\mathbf{r}$ and $\mathbf{B}_{av} = \frac{1}{V_{cell}} \int_\Omega \mathbf{B} e^{-i\mathbf{k}\cdot\mathbf{r}} d^3\mathbf{r}$ over the unit cell $\Omega$ with volume $V_{cell}$.

On the other hand, the generalized polarization vector is defined by:

$$\mathbf{P}_g = \frac{1}{-i\omega V_{cell}} \int_\Omega \mathbf{j}_d e^{-i\mathbf{k}\cdot\mathbf{r}} d^3\mathbf{r}. \tag{A2}$$

The objective of the homogenization problem is to find the effective permittivity tensor $\bar{\varepsilon}_{ef}$ such that for an arbitrary macroscopic excitation one has:

$$\bar{\varepsilon}_{ef} \cdot \mathbf{E}_{av} = \bar{\varepsilon}_h \cdot \mathbf{E}_{av} + \mathbf{P}_g. \tag{A3}$$

It is convenient to write the excitation current in terms of the macroscopic electric field. To this end, we average both sides of Eqs. (A1a) and (A1c) to obtain:

$$-\mathbf{k} \times \mathbf{E}_{av} + \omega \mathbf{B}_{av} = 0 \tag{A4a}$$

$$\omega\left(\bar{\varepsilon}_h \cdot \mathbf{E}_{av} + \mathbf{P}_g\right) + \mathbf{k} \times \frac{\mathbf{B}_{av}}{\mu_0} = -\omega \mathbf{P}_e. \tag{A4b}$$

Solving the above equations with respect to $\mathbf{P}_e$ and eliminating the magnetic field, it is found that:

$$\begin{aligned}\mathbf{P}_e &= \frac{1}{\omega^2 \mu_0}\left(k^2 \mathbf{1} - \mathbf{k}\otimes\mathbf{k} - \bar{\varepsilon}_h \omega^2 \mu_0\right)\cdot \mathbf{E}_{av} - \mathbf{P}_g \\ &\equiv \hat{\mathbf{P}}_{av}\left(\mathbf{E}_{av}\right) - \hat{\mathbf{P}}(\mathbf{E})\end{aligned} \tag{A5}$$

For convenience, we introduced the linear operators $\hat{\mathbf{P}}_{av}(\mathbf{E}_{av}) = \frac{1}{\omega^2 \mu_0}\left(k^2\mathbf{1} - \mathbf{k}\otimes\mathbf{k} - \bar{\varepsilon}_h \omega^2 \mu_0\right)\cdot\mathbf{E}_{av}$ and $\hat{\mathbf{P}}(\mathbf{E}) = \mathbf{P}_g = \frac{1}{-i\omega V_{cell}}\int_\Omega \mathbf{j}_d e^{-i\mathbf{k}\cdot\mathbf{r}} d^3\mathbf{r}$ with $\mathbf{j}_d = -i\omega(\bar{\varepsilon} - \bar{\varepsilon}_h)\cdot\mathbf{E}$. From Eq. (A5), the homogenization problem can be reformulated as an integral-differential system [64],

$$\nabla \times \mathbf{E} = i\omega \mathbf{B}, \tag{A6a}$$

$$\nabla \times \frac{\mathbf{B}}{\mu_0} = -i\omega \left[ \hat{\mathbf{P}}_{av}(\mathbf{E}_{av}) - \hat{\mathbf{P}}(\mathbf{E}) \right] e^{+i\mathbf{k}\cdot\mathbf{r}} - i\omega\overline{\varepsilon}(\mathbf{r}) \cdot \mathbf{E}. \tag{A6b}$$

where the excitation term now depends explicitly on the macroscopic electric field.

The solution of Eq. (A6) can be formally written in terms of a periodic Green's function $\mathbf{G}_p = \mathbf{G}_p(\mathbf{r},\mathbf{r}';\omega,\mathbf{k})$ that satisfies:

$$\nabla \times \nabla \times \mathbf{G}_p - \omega^2 \mu_0 \overline{\varepsilon}_h \cdot \mathbf{G}_p = \mathbf{1}_{3\times 3} e^{i\mathbf{k}\cdot(\mathbf{r}-\mathbf{r}')} \sum_{\mathbf{I}=(i_1,i_2,i_3)} \delta(\mathbf{r}-\mathbf{r}'-\mathbf{r}_\mathbf{I}). \tag{A7}$$

The Green's function determines the field radiated by a periodic array of electric dipoles in the gyrotropic host material. Here, $\mathbf{r}_\mathbf{I}$ is a generic lattice point of the direct lattice. In the spectral domain, the Green's function is given by the Fourier series:

$$\mathbf{G}_p(\mathbf{r},\mathbf{r}') = \sum_{\mathbf{J}=(j_1,j_2,j_3)} \mathbf{G}_\mathbf{J} e^{i\mathbf{k}_\mathbf{J}\cdot(\mathbf{r}-\mathbf{r}')}, \tag{A8}$$

with $\mathbf{k}_\mathbf{J} = \mathbf{k}_\mathbf{J}^0 + \mathbf{k}$, $\mathbf{k}_\mathbf{J}^0 = j_1\mathbf{b}_1 + j_2\mathbf{b}_2 + j_3\mathbf{b}_3$ is a generic point of the reciprocal lattice, and the Fourier coefficients $\mathbf{G}_\mathbf{J}$ such that:

$$\mathbf{G}_\mathbf{J}^{-1} = V_{cell}\left(k_\mathbf{J}^2 \mathbf{1} - \mathbf{k}_\mathbf{J} \otimes \mathbf{k}_\mathbf{J} - \omega^2 \mu_0 \overline{\varepsilon}_h\right). \tag{A9}$$

A straightforward analysis –similar to what is reported in Ref. [64] – shows that the solution of the homogenization problem (A6) satisfies the integral equation:

$$\mathbf{E}(\mathbf{r}) = \mathbf{E}_{av} e^{+i\mathbf{k}\cdot\mathbf{r}} + \int_\Omega d^3\mathbf{r}'\, \mathbf{G}_{p0}(\mathbf{r},\mathbf{r}') \cdot \left(i\omega\mu_0 \mathbf{j}_d(\mathbf{r}')\right), \tag{A10}$$

with $\mathbf{G}_{p0}(\mathbf{r},\mathbf{r}')$ given by:

$$\mathbf{G}_{p0}(\mathbf{r},\mathbf{r}') \equiv \mathbf{G}_p(\mathbf{r},\mathbf{r}') - \mathbf{G}_0 e^{i\mathbf{k}\cdot(\mathbf{r}-\mathbf{r}')} = \sum_{\mathbf{J}\neq 0} \mathbf{G}_\mathbf{J} e^{i\mathbf{k}_\mathbf{J}\cdot(\mathbf{r}-\mathbf{r}')}. \tag{A11}$$

Equation (A10) gives the microscopic field ($\mathbf{E}$) in terms of the unknown polarization current ($\mathbf{j}_d = -i\omega(\overline{\varepsilon} - \overline{\varepsilon}_h) \cdot \mathbf{E}$) on the inclusions and in terms of the macroscopic electric field $(\mathbf{E}_{av})$.

### B. Approximate solution of the homogenization problem

Next, we suppose that inclusions are cylindrical metallic wires with radius $r_w$ oriented along the *y*-direction. The permittivity of the inclusions is $\varepsilon_0 \varepsilon_m$. The current $\mathbf{j}_d = -i\omega(\bar{\varepsilon} - \bar{\varepsilon}_h) \cdot \mathbf{E}$ can be found by reducing Eq. (A10) to an integral equation. As a first approximation, it is assumed that the polarization current is uniform inside each metallic wire (apart from a propagation factor), so that $\dfrac{\mathbf{j}_d}{-i\omega} = C\hat{\mathbf{y}} e^{i\mathbf{k}\cdot\mathbf{r}}$, with *C* an unknown constant. The approximation is justified provided the radius of the rods is much smaller than the lattice period (thin wire approximation). Taking into account that $\dfrac{\mathbf{j}_d}{-i\omega} = (\varepsilon_0 \varepsilon_m \mathbf{1} - \bar{\varepsilon}_h) \cdot \mathbf{E}$, it follows from Eq. (A10) that for an observation point inside the metallic wire in the unit cell:

$$C(\varepsilon_0 \varepsilon_m \mathbf{1} - \bar{\varepsilon}_h)^{-1} \cdot \hat{\mathbf{y}} \approx \mathbf{E}_{av} + \omega^2 \mu_0 C \int_V d^3\mathbf{r}'\, \mathbf{G}_{p0}(\mathbf{r},\mathbf{r}') \cdot \hat{\mathbf{y}} e^{-i\mathbf{k}\cdot(\mathbf{r}-\mathbf{r}')}. \tag{A12}$$

Here, *V* is the metallic region in the unit cell. In order to obtain *C*, we calculate the inner product of both sides of the equation with $\hat{\mathbf{y}}$ and integrate both sides over *V*. This yields:

$$C\left[\hat{\mathbf{y}} \cdot \left(\varepsilon_m \mathbf{1} - \dfrac{\bar{\varepsilon}_h}{\varepsilon_0}\right)^{-1} \cdot \hat{\mathbf{y}} - \dfrac{1}{\pi r_w^2 a}\left(\dfrac{\omega}{c}\right)^2 \int_V d^3\mathbf{r} \int_V d^3\mathbf{r}' \hat{\mathbf{y}} \cdot \mathbf{G}_{p0}(\mathbf{r},\mathbf{r}') e^{-i\mathbf{k}\cdot(\mathbf{r}-\mathbf{r}')} \cdot \hat{\mathbf{y}}\right] = \varepsilon_0 \mathbf{E}_{av} \cdot \hat{\mathbf{y}}. \tag{A13}$$

Taking now into account that $\mathbf{P}_g = f_V C \hat{\mathbf{y}}$, where $f_V = \pi r_w^2 / a^2$ is the metal volume fraction, one readily concludes from Eq. (A3) that the effective permittivity tensor is given by:

$$\bar{\varepsilon}_{ef} = \bar{\varepsilon}_h + \varepsilon_0 \chi_{yy} \hat{\mathbf{y}} \otimes \hat{\mathbf{y}}, \qquad \chi_{yy} = \dfrac{1}{\hat{\mathbf{y}} \cdot \left(\varepsilon_m \mathbf{1} - \dfrac{\bar{\varepsilon}_h}{\varepsilon_0}\right)^{-1} \cdot \hat{\mathbf{y}} \dfrac{1}{f_V} - \kappa}, \tag{A14}$$

with

$$\begin{aligned}\kappa &= \left(\dfrac{\omega}{c}\right)^2 \dfrac{a}{(\pi r_w^2)^2} \int_V d^3\mathbf{r} \int_V d^3\mathbf{r}'\, \hat{\mathbf{y}} \cdot \mathbf{G}_{p0}(\mathbf{r},\mathbf{r}') e^{-i\mathbf{k}\cdot(\mathbf{r}-\mathbf{r}')} \cdot \hat{\mathbf{y}} \\ &= \left(\dfrac{\omega}{c}\right)^2 \sum_{\mathbf{J}\neq 0, j_2=0} \hat{\mathbf{y}} \cdot V_{cell} \mathbf{G}_{\mathbf{J}} \cdot \hat{\mathbf{y}} \dfrac{1}{\pi r_w^2} \int_D ds \dfrac{1}{\pi r_w^2} \int_D ds'\, e^{i\mathbf{k}_\mathbf{J}^0 \cdot (\mathbf{r}-\mathbf{r}')}.\end{aligned} \tag{A15}$$

Here $D$ is the circular cross-section of the metallic wire. We evaluate the integrals using

$$\frac{1}{\pi r_w^2}\int_D ds\, e^{i\mathbf{k}_J^0 \cdot \mathbf{r}} \approx \frac{1}{2\pi}\int_0^{2\pi} d\varphi\, e^{i\mathbf{k}_J^0 r_w \cdot (\cos\varphi, 0, \sin\varphi)} = J_0\left(\left|\mathbf{k}_J^0\right| r_w\right),$$

where $J_0$ is the zero-order Bessel function of the first kind. This yields the result:

$$\kappa = \left(\frac{\omega}{c}\right)^2 \sum_{\mathbf{J}\neq 0, j_2=0} \hat{\mathbf{y}} \cdot \mathbf{g}_\mathbf{J}(\omega, \mathbf{k}) \cdot \hat{\mathbf{y}} \left[J_0\left(\left|\mathbf{k}_J^0\right| r_w\right)\right]^2. \tag{A16}$$

where we introduced $\mathbf{g}_\mathbf{J}(\omega,\mathbf{k}) = \left(k_\mathbf{J}^2 \mathbf{1} - \mathbf{k}_\mathbf{J}\otimes\mathbf{k}_\mathbf{J} - \frac{\omega^2}{c^2}\frac{\bar{\varepsilon}_h}{\varepsilon_0}\right)^{-1}$. The series in Eq. (A16) is a double summation over $\mathbf{J} = (j_1, 0, j_3)$ with $j_1, j_3$ integers that cannot simultaneously vanish.

## C. Taylor series approximation

In the following, we evaluate $\kappa$ [Eq. (A16)] using a Taylor series in $\omega$ and $k_y$ (component of the wave vector along the metallic wires). It is implicit that

$$\mathbf{g}_\mathbf{J}(\omega,\mathbf{k}) = \left(k_\mathbf{J}^2 \mathbf{1} - \mathbf{k}_\mathbf{J}\otimes\mathbf{k}_\mathbf{J} - \frac{\omega^2}{c^2}\frac{\bar{\varepsilon}_h}{\varepsilon_0}\right)^{-1} \text{ with } \frac{\bar{\varepsilon}_h}{\varepsilon_0} = \varepsilon_t \mathbf{1}_t + i\varepsilon_g \hat{\mathbf{z}}\times\mathbf{1}_t + \varepsilon_a \hat{\mathbf{z}}\otimes\hat{\mathbf{z}}, \text{ as in the main text.}$$

The parameters $\varepsilon_t, \varepsilon_g, \varepsilon_a$ are treated as constants in the evaluation of the Taylor series. Retaining only terms up to order two, one obtains:

$$\kappa \approx \sum_{(j_1,j_3)\neq 0} \left[ \frac{\left(\frac{\omega}{c}\right)^2}{\left(\frac{2\pi}{a}j_1\right)^2 + \left(\frac{2\pi}{a}j_3\right)^2} - \frac{k_y^2}{\varepsilon_t\left(\frac{2\pi}{a}j_1\right)^2 + \varepsilon_t\left(\frac{2\pi}{a}j_3\right)^2} \right] \left[J_0\left(\left|\mathbf{k}_J^0\right| r_w\right)\right]^2. \tag{A17}$$

The above formula can be written as:

$$\kappa \approx \frac{1}{\beta_p^2}\left(\frac{\omega}{c}\right)^2 - \frac{1}{\beta_\varepsilon^2(\omega)} k_y^2. \tag{A18}$$

with $\dfrac{1}{\beta_p^2} = \sum_{\mathbf{J}\neq 0, j_2=0} \dfrac{\left[J_0\left(\left|\mathbf{k}_\mathbf{J}^0\right|r_w\right)\right]^2}{\left|\mathbf{k}_\mathbf{J}^0\right|^2}$ and $\dfrac{1}{\beta_\varepsilon^2(\omega)} = \sum_{\mathbf{J}\neq 0, j_2=0} \dfrac{\left[J_0\left(\left|\mathbf{k}_\mathbf{J}^0\right|r_w\right)\right]^2}{\varepsilon_t(\omega)\left(\dfrac{2\pi}{a}j_1\right)^2 + \varepsilon_a(\omega)\left(\dfrac{2\pi}{a}j_3\right)^2}$.

Thereby, taking also into account that $\hat{\mathbf{y}} \cdot \left(\varepsilon_m \mathbf{1} - \dfrac{\bar{\varepsilon}_h}{\varepsilon_0}\right)^{-1} \cdot \hat{\mathbf{y}} = \dfrac{\varepsilon_m - \varepsilon_t}{\left(\varepsilon_m - \varepsilon_t\right)^2 - \varepsilon_g^2}$, one finds that the

metamaterial permittivity can finally be expressed as:

$$\bar{\varepsilon}_{ef} = \bar{\varepsilon}_h + \varepsilon_0 \chi_{yy} \hat{\mathbf{y}} \otimes \hat{\mathbf{y}}, \qquad \chi_{yy} = \dfrac{1}{\dfrac{\varepsilon_m - \varepsilon_t}{\left(\varepsilon_m - \varepsilon_t\right)^2 - \varepsilon_g^2}\dfrac{1}{f_V} - \left(\dfrac{1}{\beta_p^2}\left(\dfrac{\omega}{c}\right)^2 - \dfrac{1}{\beta_\varepsilon^2(\omega)}k_y^2\right)}. \qquad (A19)$$

## Appendix B: Symmetry of the Scattering Parameters

Here we study the symmetry relations of the scattering parameters of the metamaterial slab. Let us consider different types of illumination of the metamaterial slab, as depicted in Fig. B1a. The incoming wave may arrive either from the top or from the bottom air region, and the angle of incidence is $\pm\theta$, so that it is possible to have 4 distinct propagation channels.

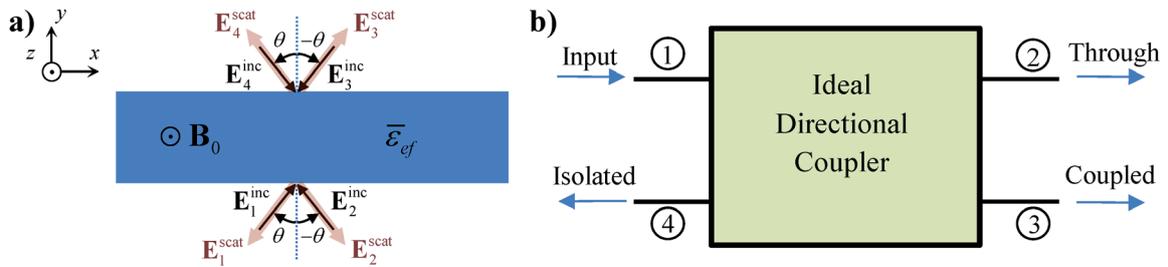

Fig. B1 **a)** Metamaterial slab illuminated by different plane waves. **b)** Equivalent 4-port network where each port is associated with a different illumination channel. The 4-port network behaves as an ideal directional coupler.

This system may be regarded as a 4-port network, where each port is associated with a specific propagation channel, as depicted in Fig. B1b. It is evident that the scattering matrix of the network satisfies:

$$\mathbf{S}(B_0) = \begin{pmatrix} 0 & S_{12} & S_{13} & 0 \\ S_{21} & 0 & 0 & S_{24} \\ S_{31} & 0 & 0 & S_{34} \\ 0 & S_{42} & S_{43} & 0 \end{pmatrix}. \tag{B1}$$

In fact, an incoming wave propagating in channel 1, only generates outgoing waves in channels 2 and 3, and so on. Moreover, taking into account the 180º-rotation symmetry of the system about the z-axis, it is possible to link the scattering parameters for excitations of ports 1 and 3, as well as the scattering parameters for excitations of ports 2 and 4:

$$\mathbf{S}(B_0) = \begin{pmatrix} 0 & R_2 & T_1 & 0 \\ R_1 & 0 & 0 & T_2 \\ T_1 & 0 & 0 & R_2 \\ 0 & T_2 & R_1 & 0 \end{pmatrix}. \tag{B2}$$

In the $j$-th matrix column the $R$ and $T$ parameters represent the reflection and transmission coefficients for an excitation of the corresponding $j$-th physical channel. The structure of the scattering matrix implies that the reflection coefficients for top and bottom illuminations and symmetric incidence angles are linked as $R_{\text{BOT}}(\pm\theta) = R_{\text{TOP}}(\mp\theta)$.

Next, we note that because of the reciprocity theorem the scattering matrix transforms $\mathbf{S} \to \mathbf{S}^T$ when the static magnetic field is flipped. Hence, we can write:

$$\mathbf{S}(-B_0) = \mathbf{S}^T(B_0) = \begin{pmatrix} 0 & R_1 & T_1 & 0 \\ R_2 & 0 & 0 & T_2 \\ T_1 & 0 & 0 & R_1 \\ 0 & T_2 & R_2 & 0 \end{pmatrix}. \tag{B3}$$

On the other hand, a 180º rotation around the $y$-axis is equivalent to interchange the roles of ports 1 and 2, and also the roles of ports 3 and 4, so that:

$$\mathbf{S}(-B_0, 180º \text{ rotation } y\text{-axis}) = \begin{pmatrix} 0 & R_2 & T_2 & 0 \\ R_1 & 0 & 0 & T_1 \\ T_2 & 0 & 0 & R_2 \\ 0 & T_1 & R_1 & 0 \end{pmatrix}. \tag{B4}$$

However, the magnetic field bias flip followed by the 180º rotation around the *y*-axis leads to the original configuration. This implies that $\mathbf{S}(+B_0) = \mathbf{S}(-B_0, 180º \text{ rotation y-axis})$, or equivalently that $T_1 = T_2$. Hence, we finally conclude that $T_{\text{BOT}}(\theta) = T_{\text{BOT}}(-\theta) = T_{\text{TOP}}(\theta) = T_{\text{TOP}}(-\theta)$, i.e., the transmission coefficient is independent of the illumination channel.